# A High-Throughput Multi-Mode LDPC Decoder for 5G NR

Sina Pourjabar, Gwan S. Choi
Department of Electrical and Computer Engineering Texas A&M University, College Station, TX
Email: {pourjabar, gwanchoi}@tamu.edu

*Abstract*— This paper presents a partially parallel low-density parity-check (LDPC) decoder designed for the 5G New Radio (NR) standard. The design is using a multi-block parallel architecture with a flooding schedule. The decoder can support any code rates and code lengths up to the lifting size $Z_{max}$= 96. To compensate for the dropped throughput associated with the smaller Z values, the design can double and quadruple its parallelism when lifting sizes $Z \leq 48$ and $Z \leq 24$ are selected respectively. Therefore, the decoder can process up to eight frames and restore the throughput to the maximum. To facilitate the design's architecture, a new variable node for decoding the extended parity bits existing in the lower code rates is proposed. The FPGA implementation of the decoder results in a throughput of 2.1 Gbps decoding the 11/12 code rate. Additionally, the synthesized decoder using the 28 nm TSMC technology, achieves a maximum clock frequency of 526 MHz and a throughput of 13.46 Gbps. The core decoder occupies 1.03 mm$^2$, and the power consumption is 229 mW.

## I. INTRODUCTION

With the introduction of 5G and its wide range of supported code rates and lifting values, having a reasonable size and flexible decoder is demanding. Two types of forward error correction codes are deployed in 5G [1]. Polar codes are used for decoding the control channel, and low-density parity check (LDPC) codes [2] are used for decoding the data channel. Compared to the previously defined standards, such as IEEE 802.16e and IEEE 802.11ad, where each code rate has its own parity matrix, the base graph in 5G can be used to implement different code rates. There are two base graphs (BG) defined for 5G, BG1, and BG2 [3]. BG1 is a 46×68 matrix where the mother code is an embedded 4×26 sub-matrix. The first 22 columns in BG1 include the information part of the base parity matrix. The lifting size Z can be selected from sets of values defined by the standard with the minimum of Z=2 and maximum of Z=384 [4]. On the encoder side, the first 2×Z columns containing the information part are always punctured, and the rest of the data is transmitted. The mother code in BG1 yields the highest code rate of 11/12. The extended section of the matrix provides additional parity bits to improve the decoding performance in higher channel signal-to-noise ratio (SNR) conditions while yielding a lower code rate. The lower limit code rate for BG1 is 1/3. On the other hand, BG2 is a 42×52 matrix in which the information part is included in the first ten columns. BG2's code rate can vary between 1/5 and 2/3. The decoding strategy in this work is implemented on the BG1. However, because of similarities between BG1 and BG2, the same approach can be applied for decoding BG2.

Flooding and layered decoding are the most common schedules used in LDPC decoders [5]. In BG1, since the first four layers contain the information, they always need to be decoded. Compared to other layers, the first four layers have more common nodes, and thus data dependency between them is more evident. Consequently, in a multi-block parallel architecture, increasing the parallelism of the blocks makes the layered decoding more challenging to implement. This occurs because managing memory access times becomes a difficult task, and if not properly scheduled, it can lead to an increased number of stalls. For this reason, in this work, flooding is favored over layered decoding.

In a partially parallel architecture, the shift network is responsible for connecting the variable nodes (VN) to the check nodes (CN) in the correct order. The shift network has a key role in defining the reconfigurability, parallelism, area, and routing delay of a decoder. Although introducing flexibility to a decoder for supporting different code lengths is desirable, it also increases the shift network's area and requires a more complex control unit for generating shift signals. In such designs, regardless of the Z value, the same number of clock cycles per iteration is required. Since throughput has a direct relation with the lifting size, selecting a smaller Z value will drop the decoder's throughput. Moreover, since the number of 2×2 switches is initially defined to support a $Z_{max} \times Z_{max}$ input, Z values smaller than $Z_{max}$ will face the same number of switching stages. This results in a longer routing network compared to a design that is purposefully tailored to support a single smaller Z value. 5G code length is larger than the previously defined wireless standards and supports a wider range of code rates and code lengths. So, we need to keep a reasonable balance between the hardware area, throughput, power consumption, and the flexibility of the decoder.

In this paper, a multi-block parallel decoder specific to the 5G parity matrix is proposed. For lifting values smaller than $Z_{max}/2$ the decoder provides two decoding strategies. In the first option, similar to the conventional decoders, the unused CNs and VNs are disabled to save power. This will also result in a lower throughput compared to when Z=$Z_{max}$. In the second option, depending on the code length, the decoder can increase its parallelism and reuse the remaining CNs and VNs to process additional frames and maintain the throughput at maximum.

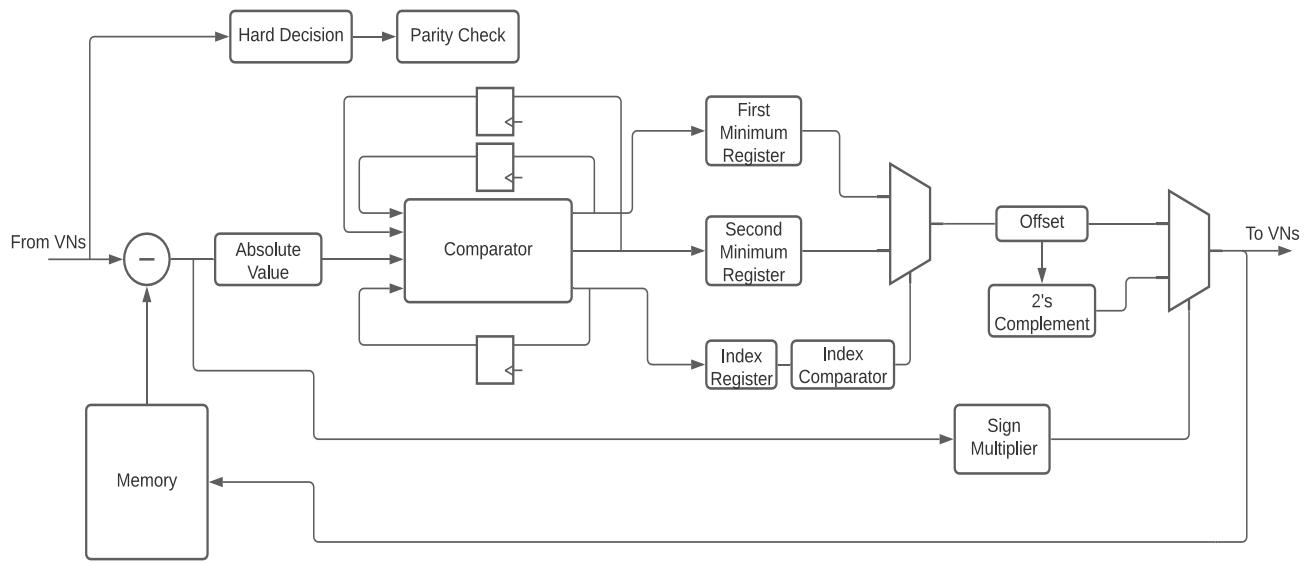

Fig. 1. Check node architecture with early decoding termination.

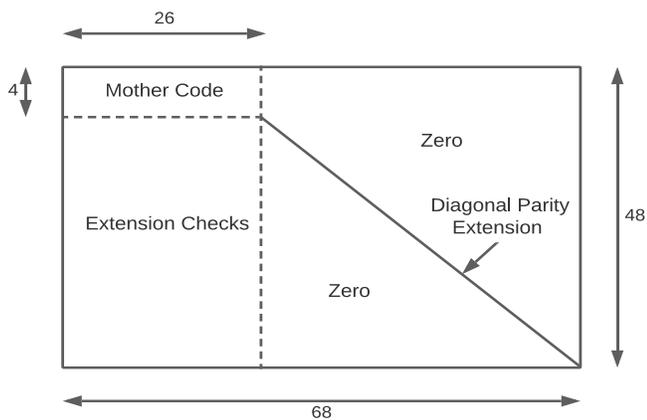

Fig. 2. 5G parity matrix structure for the base graph 1.

The design also introduces a smaller variable node unit specific to processing the diagonal parity extension for lower code rates in 5G.

The rest of the paper is organized as follows. Section II explains the decoding strategy based on the BG1 parity matrix. It also describes the CN and VN architectures. A simplified architecture for decoding the extended VNs is also proposed. Section III explains the shift network architecture supporting any shift value and any Z size up to $Z_{max}=96$. Additionally, it explains how the shift network can help increase the design's parallelism and compensate for the low throughput when smaller Z values are selected. Section IV discusses the top-level decoder and the FPGA implementation using Virtex 7 and post-synthesis using the 28 nm technology. The design is also compared with other works. Finally, section V concludes the paper.

## II. ARCHITECTURE

### A. Check Node

The CN supports up to 16 inputs and is optimized for the BG1 parity matrix. Evaluating the connection graph in Fig. 2 shows that the first 26 columns always participate in decoding. Compared to IEEE 802.16e and IEEE 802.11ad, the base parity matrix size in 5G is large. Therefore, doing a fully parallel VN implementation in a row-centric structure will be inefficient in terms of hardware utilization. Hence, at each clock cycle, $13 \times Z$ VN messages equal to a half layer are loaded into the CNs. Also, decoding the extended parity bits requires additional Z VN messages to be loaded into the CNs.

The decoder is implemented in a multi-block parallel fashion. In Fig. 1, a CN receives each layer's messages from the VNs in two clock cycles. The input memory holds the previous iteration messages sent from a CN to a VN. Before the subtraction of new data from the old data, the hard decision result is checked for error detection and the possibility of the early decoding termination if no errors were found. If there are errors in the data and the maximum number of iterations have not been reached, the decoding will continue. The 16-input comparator is implemented using the tree structure algorithm [6] and works in a serial manner. Initially, it receives the first half layer data to find the first and second minimum values as well as the index for the first minimum value. Afterward, the results are sent back into the comparator input to be included in the second half layer comparison. The final minimum values for each layer are stored in the minimum values registers and the index register. After storing the minimum values for all layers, the second stage of the CN starts to apply offset and compensate for the overestimation of the Min-Sum algorithm. Subsequently, the updated values are directed to the VNs as well as the input memory in the CN. As shown in Fig. 1, to avoid stalls inside the CN, two frames are processed at a time. While the incoming frame's minimum values are being processed, the outgoing frame values are sign multiplied and sent to the VNs.

### B. Variable Node

Two types of VNs are deployed for the 5G decoder. The primary VNs shown in Fig. 3(a) have two registers. At the beginning of each iteration, the intrinsic channel information and the updated messages from the CNs are added together. In case of an overflow, the output will saturate to a pre-defined value. The summation result is stored in the incoming frame register. A feedback loop helps to add each layer's data with the next layers accordingly. Once the required number of layers are processed, the accumulation results are forwarded into the outgoing frame register to be loaded into the CNs in the next clock cycles.

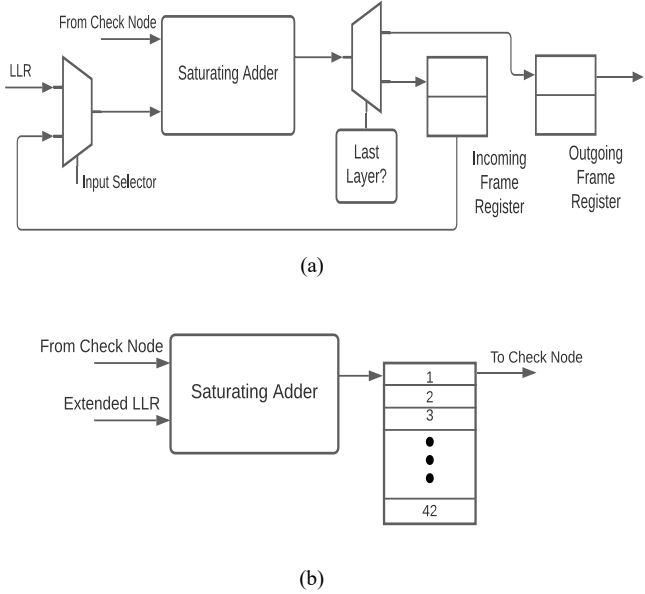

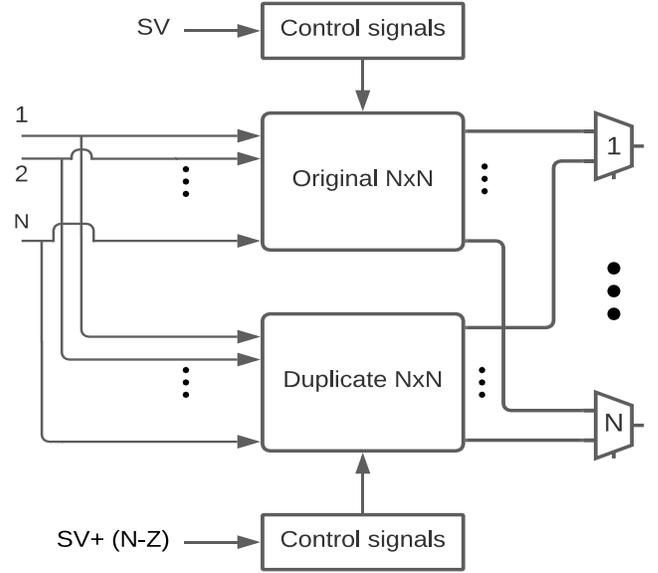

Fig. 3. (a) Primary variable node architecture for decoding the mother code.
(b) Secondary variable node for decoding the extended parity bits.

Fig. 4. N×N Banyan variant network. Shift value (SV) ≤ active inputs (Z)≤ N [10].

The second type of VN novel in this work is depicted in Fig. 3(b). The extended VNs (EVN) are utilized for decoding columns 27 through 68 required for lower code rates. The extended section of the parity matrix in 5G shows that each EVN has only one CN connection, and therefore, there is no need for the accumulation register or the feedback loop. This will also result in a more straightforward controller unit for the EVN. The EVN's architecture can be simplified to an adder and a single memory unit with the depth of 42 to hold messages from columns 27 to 68. Therefore, there is a simultaneous read and write from the memory for two different frames. Z EVNs process one column at each clock cycle. A single EVN utilizes 16 LUTs and six flip-flops in an FPGA implementation, while a VN utilizes 38 LUTs and 36 flip-flops. This VN simplification is unique to 5G. Because unlike 5G, IEEE802.11n and IEEE 802.11ad have different base parity matrices for each code rate, and therefore a comprehensive simplification of parity VN applicable to all code rates is not possible. The inclusion of a single diagonal extension in the 5G parity matrix shown in Fig. 2 helps create a unified parity matrix that supports all code rates. This inherent compatibility also benefits the decoding process in terms of reducing the initial decoding latency [7]. The decoder can start decoding as soon as it receives the mother code, which is the shortest code length.

C. Shift Network

Selecting an appropriate shift network has a significant role in reducing the path delay between the CNs and VNs. It can also help promote the flexibility of the decoder. Benes [8], Banyan and QSN [9] are the common networks used for shifting the messages cyclically. The Benes network has a non-blocking property, meaning N inputs can be mapped to any of the N different outputs. The network consists of 2×2 switches where each switch is made of two multiplexers in parallel. An N×N Benes network contains $2N \times \log_2(N) - N$ multiplexers in which $2 \times \log_2(N) - 1$ of the multiplexers make up the blocking property. This means that in some of the arbitrary mappings, some inputs may need to access the same switch output in their path from input to the final output. However, in a cyclically shifting scheme, blocking does not occur for the Banyan network. Compared to the Benes network, the Banyan network has a shorter critical path consisting of $\log_2(N)$ stages and can be implemented using $N \times \log_2 N$ multiplexers.

QSN network is another type of shifting network in which the shift network itself is divided into a left shift, a right shift and a merge network. QSN also has a shorter critical path made of $\log_2(N) + 1$ stages compared to the Benes network. An N×N Banyan network with Z enabled input, can perform cyclic shift only if all inputs are utilized (Z=N). Whereas an N×N QSN network can still perform cyclic shifts when Z≤ N and shift value (SV) ≤ Z.

If a decoder is targeted for one code length, because of having a smaller footprint and less complexity, the Banyan network is the preferable option compared to QSN. However, for supporting multiple code lengths and further reconfigurability, QSN network is required. [10] introduced a variant of the Banyan network, that similar to the QSN network can support cyclical shifts for Z ≤N. As shown in Fig. 4, in this network, a duplicate of the original Banyan network and an additional multiplexer stage are added to the design. For an N×N switch with Z active inputs and shift value SV, the original network shifts inputs by SV, whereas the duplicate network shifts by SV+ (N-Z). Basically, each network can shift a portion of the active inputs correctly, and in the last stage, the multiplexers select the correct shift results between the two networks.

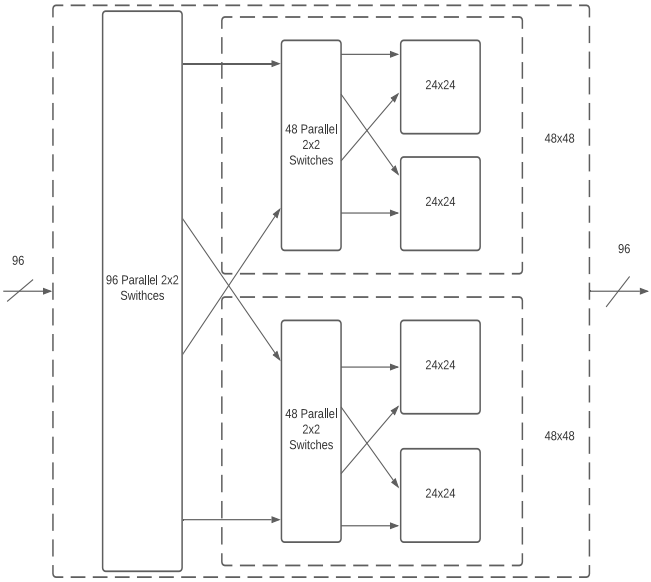

Fig. 5. A 96×96 shift network. When required each sub network can work as an independent shift network.

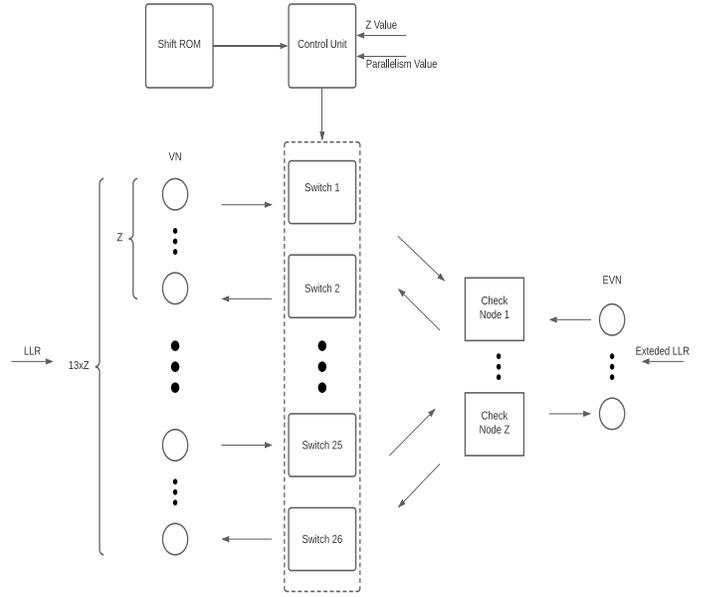

Fig. 6. Partially parallel architecture of the proposed 5G decoder.

TABLE I
FPGA RESULTS COMPARISON

|  | Proposed | [11] | [12] | [13] |
|---|---|---|---|---|
| Code Length | 2496-6528 | 2304 | 648-1944 | 26112 |
| Standard | 5G NR | IEEE 802.16e (WiMAX) | IEEE802.11n (Wi-Fi) | 5G NR |
| Decoding Algorithm | Offset Min-Sum | Min-Sum | Offset Min-Sum | Offset Min-Sum |
| Architecture | Partially Parallel | Partially Parallel | Partially Parallel | Fully Parallel |
| Frequency (MHz) | 82 | 260 | 116 | 102 |
| Quantization (Bits) | 5 | 6 | 6 | 7 |
| Maximum Iterations | 10 | 20 | 10 | 10 |
| Throughput (Mbps) | 490.3-2168 | 290 | 617-1808 | 2900 |
| LUT | 225191 | - | 67204 | 1448762 |
| Block RAM | 96 | 38 | - | - |
| Decoding Schedule | Flooding | Flooding | Layered | Layered |
| FPGA | Virtex 7- XC7VX690T | Virtex 7-VX485T | Virtex4- XC4vlx160 | Virtex Ultrascale+ |

Network structures proposed in [9] and [10] can support any expansion factor $Z \leq N = Z_{max}$ and have a similar critical path. The QSN network uses fewer multiplexers than the Banyan variant. However, because of their additional reconfigurability, both networks still have large footprints. The Banyan variant network has an additional benefit that makes it suitable for designs requiring high throughput. In a block parallel decoder supporting different code lengths, selecting a lifting size smaller than $Z_{max}$ translates into not using part of the shift network and disabling $(Z_{max}-Z)$ CNs. This will also affect the decoder's throughput. A decoder's throughput, regardless of its architecture, is defined by:

Throughput = Decoded frames per second × Base matrix size × Z.    (1)

As equation (1) shows, smaller Z values reduce the throughput of the decoder. An advantage of the Banyan variant over the QSN is that the shift network in the Banyan variant can be reconfigured to work as multiple independent smaller networks. This way, if $Z_{max}/4 < Z \leq Z_{max}/2$, each $Z_{max} \times Z_{max}$ network can perform as two smaller $Z_{max}/2 \times Z_{max}/2$ shift networks. Therefore, each smaller network can process a different frame, and the total throughput is doubled. As Fig. 5 illustrates, for a shift network of $Z_{max}=96$, two Z=48 can output the equal amount of throughput. Likewise, any Z between 25 and 48 can use two independent decoders, each decoding a different frame. Applying the same approach, for $Z \leq Z_{max}/4$, four smaller networks can be utilized.

In our proposed decoder, this switching technique is incorporated to increase the parallelism and reuse the unused part of the shift network and the disabled CNs and VNs to restore the decoding throughput to the maximum.

III. DECODER IMPLEMENTATION RESULTS

As shown in Fig. 6, there are two types of VNs in the design. The majority of the VNs have the architecture shown in Fig. 3(a) and are connected to the CNs via a shift network. Decoding lower code rates requires extended Log-likelihood ratios (LLRs) to be loaded into the EVNs. Each EVN has only one connection to a CN. The 5G parity shift sets show that the cyclic shift value between the EVNs and the CNs is equal to zero. Therefore, EVNs can connect to CNs without the need for a shift network. These VNs are only enabled for decoding a lower rate code. When there is no connection between a CN and a VN, the CN comparator's input will be saturated. Subsequently, the unconnected VN will be disabled for reading and writing.

TABLE II
SYNTHESIS RESULTS AND PERFORMANCE SUMMARY

|  | Proposed | [14] | [15] | [16] |
|---|---|---|---|---|
| Technology(nm) | 28 | 28 | 40 | 90 |
| Standard | 5G NR | IEEE 802.11ad | IEEE 802.11ad | IEEE 802.11n/ac |
| Maximum Code Length | 6528 | 672 | 672 | 1944 |
| Supported Lifting Values (Z) | $Z \leq 96$ | Z=42 | Z=42 | Z=27, 54, 81 |
| Decoding Schedule | Flooding | Flooding | Flooding | Layered |
| Quantization (Bits) | 5 | 5 | 5 | 4 |
| Maximum Iterations | 10 | 10 | 7 | 10 |
| Pipeline Stages | 4 | 8 | 3 | 5 |
| Frequency (MHz) | 526 | 202 | 220 | 555 |
| Throughput (Gbps) | 13.46 | 6.78 | 6.16 | 4.5 |
| Core Area (mm2) | 1.03 | 1.99 | 0.8 | 4.88 |
| Voltage (V) | 1 | 0.9 | 1.1 | - |
| Power (mW) | 229 | 104 | 203 | 523 |
| Energy Efficiency (pJ/bit) | 17.01 | 15.34 | 32.95 | 116 |
| Implementation | Synthesis | Fabricated | Fabricated | Place and route |
| Decoding Algorithm | Offset Min-Sum | Min-Sum | Min-Sum | Modified Min-Sum |

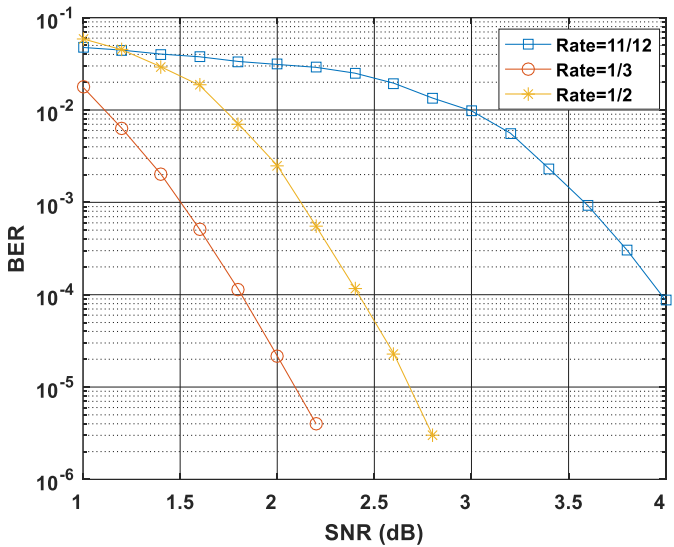

Fig. 7. BER comparison for Z=96 and selected rates 11/12, 1/2 and 1/3. Decoding performed using Offset Min-Sum algorithm with the offset value=0.5, 5- bit quantization and 10 iterations.

The proposed architecture is implemented in Xilinx Virtex 7 FPGA. Messages throughout the VN, CN, and the shift network are quantized to 5 bits. The shift network path consists of seven stages, and to further increase the throughput of the decoder, the output of the network is pipelined. The critical path of design is between the first memory unit in the CN and the output of the comparator unit. In our FPGA simulation, this memory is implemented using block RAMs, which helps reduce the number of LUTs used as memory. Table I compares the proposed work with the other reported architectures. Decoding the code rate 11/12 requires 18 clock cycles for each iteration. With the maximum number of iterations set to 10 and a clock frequency of 82 MHz, a throughput of 2.1 Gbps is achievable. Fig. 7, compares the bit error rate (BER) performance of multiple code rates assuming binary phase-shift keying (BPSK) modulation over an additive white Gaussian noise (AWGN) channel. To provide a better estimate of the area and the power consumption, the decoder is also synthesized using the available TSMC 28 nm technology. A comparison of the result is presented in Table II.

In this section, the parallelism and the flexibility of our decoder are compared to other works. The decoder in [11] uses a barrel shifter that only supports Z=96 and half-rate WiMAX (1152, 2304). In [12], each sub-block in the parity matrix of IEEE 802.11n is shifted using a barrel shifter and a RAM. The Barrel shifter input size is equal to the minimum supported lifting value ($Z_{min}$=27) to help minimize the shift network area. When $Z=Z_{min}$, cyclic shifts happen in one clock cycle. For Z=54 and $Z=Z_{max}$=81, the shift network uses a RAM to store the shifted values. Later, an address generator helps read the messages in the correct order from the RAM. As mentioned in [12], this process creates latencies of two and three clock cycles in the shift network for Z=54 and Z=81, respectively. In [14] and [15], the parity matrix of IEEE 802.11ad only supports Z=42 and the code length of 672 bits. This is an inherent advantage of the parity matrix. As a result, the decoder requires a smaller and less complex shift network. Consequently, a shorter critical path and a higher throughput can be achieved. The decoder in [16] supports all three lifting values and four code rates defined in the IEEE802.11n standard. In this design for transferring messages from VN to CN and from CN to VN, only one shift network is used. This results in a smaller shift network area. However, since CN and VN can't access the shift network simultaneously, additional stalls are introduced into the design when decoding multiple frames concurrently. Decoding codes with smaller Z values will also reduce the total throughput since the shift network doesn't support multi-frame parallelism.

Compared to the other works, our proposed 5G decoder can act as multiple separate decoders when $Z \leq Z_{max}/2$ and make the overall throughput less dependent on the lifting value.

IV. CONCLUSION

This paper proposed a five-stage pipelined multi-mode LDPC decoder for the 5G NR standard. The design is capable of decoding all the code rates supported by the standard and any lifting values up to $Z_{max}=96$. Deploying the flooding schedule, a simplified VN for decoding the extended parity bits for the lower code rates is introduced. Additionally, the proposed decoder helps minimize the effect of selecting smaller Z values on reducing the throughput. For $Z \leq Z_{max}/2$ and $Z \leq Z_{max}/4$, the decoder can reconfigure the shift network and reuse the unutilized resources to increase the number of processed frames and improve the total throughput.